# Acoustic Willis metamaterials beyond the passivity bound

Choonlae Cho[1,2†], Xinhua Wen[1†], Namkyoo Park[2]*, Jensen Li[1]*

[1] *Department of Physics, The Hong Kong University of Science and Technology, Clear Water Bay, Hong Kong, China.*

[2] *Photonic Systems Laboratory, Department of Electrical and Computer Engineering, Seoul National University, Seoul, Korea 08826*

† Equal contributions
*E-mail address for correspondence: jensenli@ust.hk, nkpark@snu.ac.kr

**Acoustic bianisotropy, also known as the Willis parameter, expands the field of acoustics by providing nonconventional couplings between momentum and strain in constitutive relations. Sharing the common ground with electromagnetics, the realization of acoustic bianisotropy enables the exotic manipulation of acoustic waves in cooperation with a properly designed inverse bulk modulus and mass density. While the control of entire constitutive parameters substantiates intriguing theoretical and practical applications, a Willis metamaterial that enables independently and precisely designed polarizabilities has yet to be developed to overcome the present restrictions of the maximum Willis bound and the nonreciprocity inherent to the passivity of metamaterials. Here, by extending the recently developed concept of virtualized metamaterials, we propose acoustic Willis metamaterials that break the passivity and reciprocity limit while also achieving decoupled control of all constitutive parameters with designed frequency responses. By instituting basis convolution kernels based on parity symmetry for each polarization response, we experimentally demonstrate bianisotropy beyond the limit of passive media. Furthermore, based on the notion of inverse design of the frequency dispersion by means of digital convolution, purely nonreciprocal media and media with a broadband, flat-response Willis coupling are also demonstrated. Our approach offers all possible independently programmable extreme constitutive parameters and frequency dispersion tunability accessible within the causality condition and provides a flexible platform for realizing the full capabilities of acoustic metamaterials.**

**Introduction**

Metamaterials exhibit various extraordinary wave phenomena and enable novel applications by providing unconventional wave properties in many wave systems[1-3]. In acoustic systems, unusual wave parameters such as extremely high[4], negative[5,6], or zero bulk moduli and mass densities[7,8] have now also become reality, enabling exotic applications such as superfocusing[9,10], extraordinary diffraction[11], and acoustic cloaking[12-14]. Recently, with the experimental realization of the Willis coupling parameter[15] as an acoustic duality of bianisotropy in electromagnetic waves, the scope and capability of acoustic metamaterials have been greatly extended by allowing coupling between pressure and velocity fields[16]. Sharing the same design principle of symmetry breaking used in omega media or moving media in electromagnetics[17], Willis metamaterials have been realized theoretically and experimentally on various platforms in recent years[7,18-22]. Combined with a properly designed inverse bulk modulus and mass density, Willis couplings now enable applications that were impossible before, such as the separate control of reflected and transmitted waves[6] and diffraction-free metasurfaces[23] signifying the further generalization of the generalized Snell's law, bianisotropic nihility[24], and metagratings[25].

Despite the great successes achieved thus far, the full potential of Willis metamaterials has not been achieved. The maximum bound of Willis coupling[25] and nonreciprocal operation, which are inherent to the passivity of the metamaterial structure, currently hinder the full exploitation of the advantages offered by Willis metamaterials in future applications. Although the breaking of the passive Willis bound or the tuning of nonreciprocity have been envisaged with the introduction of active metamaterials[26], the question of how to achieve selective excitation and flexible control of all four constitutive parameters to enable extreme bianisotropy and fully controllable nonreciprocity has yet to be answered.

In this work, we assess the ultimate bound of Willis coupling and the conditions for nonreciprocity under the introduction of activity and implement the idea based on the recently developed concept of virtualized meta-atoms[27], which replace the scattering functions of physical metamaterials with deliberately designed software convolution functions. By achieving selective excitation in addition to precise balancing between cross-coupling terms, the inverse bulk modulus, and the mass density, we demonstrate full nonreciprocity and the ultimate Willis bound within the same platform as a universal building block for future Willis applications. As an example of the wide-open flexibility of the proposed concept, we further realize Willis atom operation with broadband and flat bianisotropy, in both the purely reciprocal and purely nonreciprocal regimes, from the dispersion curve analytically constructed via the inverse design method.

**Results**

**Polarization process based on parity symmetry.** We define a scattering matrix $\mathbf{S}$ having parity symmetry[21] in the one-dimensional system as depicted in Fig. 1, for incident ($a$) and scattered ($b$) waves propagating in the forward (+) and backward (–) directions, which are decomposed into components of even ($e$) and odd ($o$) parity: $a_e = (a_+ + a_-)/2$, $a_o = (a_+ - a_-)/2$, $b_e = (b_+ + b_-)/2$, and $b_o = (b_+ - b_-)/2$. The couplings between the incident fields and scattered fields are then written as $(b_e\ b_o)^T = \mathbf{S}\ (a_e\ a_o)^T$, with the scattering matrix $\mathbf{S}$ being defined as

$$\begin{pmatrix} s_{ee} & s_{eo} \\ s_{oe} & s_{oo} \end{pmatrix} = \frac{1}{2}\begin{pmatrix} t_+ + t_- + r_+ + r_- - 2 & t_+ - t_- + r_+ - r_- \\ t_+ - t_- - r_+ + r_- & t_+ + t_- - r_+ - r_- - 2 \end{pmatrix}, \quad (1)$$

where $r$ and $t$ are the reflection and transmission coefficients, respectively. Then the relation between the scattering matrix $\mathbf{S}$ and the normalized polarizability $\boldsymbol{\alpha}$ becomes:

$$\begin{pmatrix} \alpha_{pp} & \alpha_{pv} \\ \alpha_{vp} & \alpha_{vv} \end{pmatrix} = \frac{1}{ik_0}\begin{pmatrix} s_{ee} & s_{eo} \\ s_{oe} & s_{oo} \end{pmatrix}, \quad (2)$$

where $k_0$ is the free space wavenumber (see Supplementary Note S1 for the derivation of and detailed expressions for each element). In this representation, the diagonal terms relating even-incident to even-scattering components ($s_{ee}$) and odd-incident to odd-scattering components ($s_{oo}$) correspond to the inverse bulk modulus and mass density, respectively, while the off-diagonal components coupling even to odd components ($s_{oe}$) and odd to even components ($s_{eo}$) are the acoustic bianisotropy or Willis coupling parameters. Although our consideration of Willis coupling does not assume passivity and reciprocity in general, it is worth mentioning that for a conventional passive and reciprocal metamaterial, reciprocity is equivalently represented by $t_+ = t_-$ or $s_{eo} = -s_{oe}$, and the maximum Willis coupling[20] corresponds to $|s_{eo}|$ (or $|s_{oe}|$) = 1, which is difficult to achieve while keeping the other components intact.

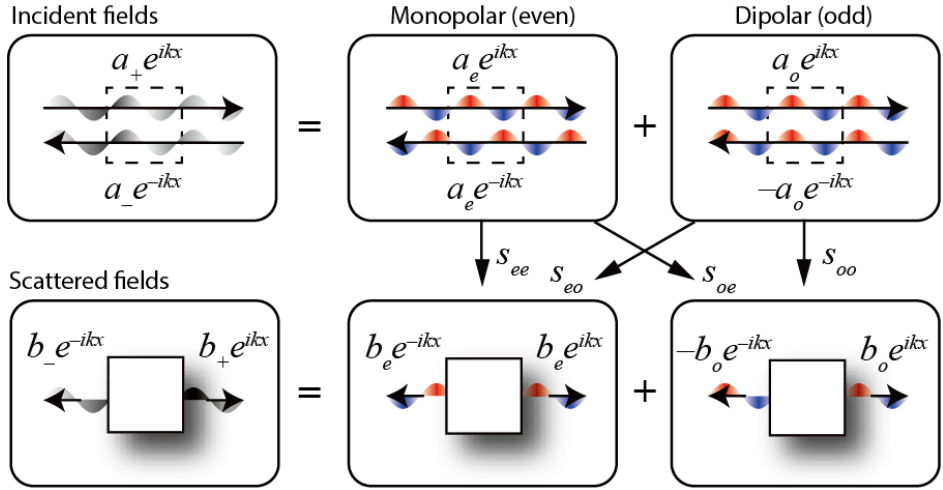

**Figure 1 | Scattering matrix S based on coupling symmetry.** The definition of the scattering matrix **S** in a one-dimensional system. The incident and resultant scattered fields propagating in the forward and backward directions are decomposed into even and odd components. The even-to-even and odd-to-odd scattering parameters correspond to the inverse bulk modulus and density, respectively, while the even-to-odd and odd-to-even scattering parameters are the Willis parameters.

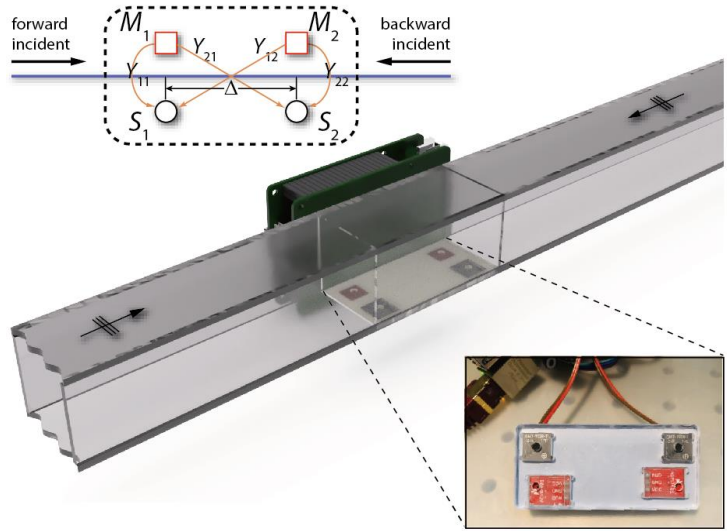

**Figure 2 | Bianisotropic virtualized metamaterial.** A virtualized metamaterial consisting of two microphones and two speakers connected to a microprocessor embedded in the cover of an acoustic waveguide. The bottom right inset shows a photograph of a transducer module of a virtualized meta-atom. The top inset shows an analytical representation of the virtualized metamaterial atom: two microphones ($M_j$) are convolved with a 2 × 2 matrix ($Y_{ij}$) returning signals to fire at the two speakers ($S_i$) as secondary radiation. The distances between the two speakers and microphones are equally set to $\Delta = 50$ mm. The scattering property of the meta-atom is tested by incident waves coming from the forward and backward directions to identify all 4 scattering parameters ($r_+$, $t_+$, $r_-$, and $t_-$).

To realize all these polarization responses with selective excitation and precise balancing between the cross-coupling terms, inverse bulk modulus, and mass density, we employ a platform of a virtualized meta-atom, which can directly mold the above parameters with the designer convolution function connecting the detectors and sources[27]. As depicted in Fig. 2, the microprocessor returns output values to two speakers ($S_i$) from the detected signals of two microphones ($M_j$) by means of the programmed convolution kernels ($\tilde{Y}_{ij}$). I.e., the output voltages of the sources are calculated in the time domain as follows:

$$S_i(t) = -\partial_t^2(\tilde{Y}_{ij}(t) * M_j(t)), \tag{3}$$

where * is the convolution operator, and the subscripts $i, j = 1$ or 2 are the labels of the speakers and microphones. In Eq. (3), one derivative is given by the software for a zero averaged offset value, and the other derivative appears in speakers when generating a pressure field in time differential way. In the frequency domain, entire operation is summarized as $S_i(\omega) = Y_{ij}(\omega)M_j(\omega)$ where $Y_{ij}(\omega) = \omega^2\tilde{Y}_{ij}(\omega)$. To achieve a connection between the speaker output $S_i$ and the microphone-detected signal $M_j$, similar to the polarization process in Eq. (1), we decompose the convolution kernel $Y_{ij}$ by introducing a basis of convolution matrices:

$$\mathbf{e}_{ee} = \frac{1}{2}\begin{pmatrix} 1 & 1 \\ 1 & 1 \end{pmatrix}, \mathbf{e}_{eo} = \frac{1}{2}\begin{pmatrix} -1 & 1 \\ -1 & 1 \end{pmatrix}, \mathbf{e}_{oe} = \frac{1}{2}\begin{pmatrix} -1 & -1 \\ 1 & 1 \end{pmatrix}, \mathbf{e}_{oo} = \frac{1}{2}\begin{pmatrix} 1 & -1 \\ -1 & 1 \end{pmatrix}, \tag{4}$$

which satisfy $\mathbf{Y} = Y_{ee}\mathbf{e}_{ee} + Y_{eo}\mathbf{e}_{eo} + Y_{oe}\mathbf{e}_{oe} + Y_{oo}\mathbf{e}_{oo}$. Eq. (3) can then be rewritten in terms of the symmetric (even) and antisymmetric (odd) components of the speakers and microphones as follows:

$$\begin{pmatrix} S_1 + S_2 \\ S_2 - S_1 \end{pmatrix} = \begin{pmatrix} Y_{ee} & Y_{eo} \\ Y_{oe} & Y_{oo} \end{pmatrix}\begin{pmatrix} M_1 + M_2 \\ M_2 - M_1 \end{pmatrix}. \tag{5}$$

It is important to note that in this representation in terms of a basis of convolution matrices, each basis matrix $\mathbf{e}_{kl}$ exclusively addresses one of the four polarizability parameters, including the two Willis parameters. In our atom configuration shown in Fig. 2, in the case of selective excitation (i.e., $\mathbf{Y} = Y_0\mathbf{e}_{ij}$), the polarizability parameters can then be written as follows (see Supplementary Note S2):

$$\alpha_{pp} = \frac{4}{ik_0} \frac{\cos(k_0\Delta/2)^2 Y_0}{1-(1+e^{ik_0\Delta})Y_0},$$
$$\alpha_{pv} = 2k_0^{-1}\sin(k_0\Delta)Y_0,$$
$$\alpha_{vp} = -2k_0^{-1}\sin(k_0\Delta)Y_0, \quad (6)$$
$$\alpha_{vv} = \frac{4}{ik_0} \frac{\sin(k_0\Delta/2)^2 Y_0}{1-(1-e^{ik_0\Delta})Y_0}.$$

where $\Delta$ is the distance between the two speakers (and microphones). Therefore, decoupled control or balancing among all acoustic wave parameters can be realized with analytically constructed kernels $Y_{ij}$ of the desired design. It is emphasized that the relationship derived in Eq. (6) is analogous to the effective medium theory expression[28] that relates the constitutive parameters to the scattering parameters of composite scatterers.

Figure 3 shows an experimental demonstration of the selective excitation of each polarizability parameter in the virtual Willis metamaterial. We set the program to have one of the basis convolution matrices $\mathbf{e}_{ee}$, $\mathbf{e}_{eo}$, $\mathbf{e}_{oe}$ and $\mathbf{e}_{oo}$ given in Eq. (4) with the Lorentzian-form coefficient $\tilde{Y}_0(\omega)$. For time-domain microprocessor signal processing, $\tilde{Y}_0(\omega)$ can then be expressed or implemented as the following impulse response function:

$$\tilde{y}_0(t) = \frac{a}{\omega_0^2}\sin(\omega_0 t + \theta)e^{-\gamma t}u(t), \quad (7)$$

where $u(t)$ is the Heaviside step function, $a = 2\pi \times 15$ is the total scaling factor, $\omega_0 = 2\pi \times 1.0$ kHz is the resonance frequency, $\theta = -\pi/2$ is the phase, and $\gamma = 2\pi \times 15$ Hz is the resonance bandwidth. Figures 3a and 3b show the Lorentzian polarizations experimentally realized with even-to-even $\mathbf{e}_{ee}$ and odd-to-odd $\mathbf{e}_{oo}$ excitations, which are responsible for inverse bulk modulus and mass density, respectively, and Fig. 3c,d show the implementation of bianisotropy achieved with odd-to-even $\mathbf{e}_{eo}$ and even-to-odd $\mathbf{e}_{oe}$ convolutions. Each polarization component is exclusively excited, with the other components suppressed, in excellent agreement with the analytical results in Eq. (6). Since this approach enables simultaneous independent control of the four wave parameters, by balancing the even-to-odd and odd-to-even couplings, we can also easily realize the *purely* reciprocal and nonreciprocal Willis parameters $\kappa = i(\alpha_{pv} - \alpha_{vp})/2$ and $\chi = (\alpha_{pv} + \alpha_{vp})/2$: from the symmetric convolution kernels $\mathbf{Y} = Y_0(\mathbf{e}_{eo} + \mathbf{e}_{oe})$ for the reciprocal case (Fig. 3e) and from the antisymmetric $\mathbf{Y} = Y_0(\mathbf{e}_{eo} - \mathbf{e}_{oe})$ in the nonreciprocal case (Fig. 3f). Our results are not subject to the strict restriction imposed by the geometry of the scatterers in physical metamaterials, in contrast with previous approaches, in which the resonance strengths and bandwidths of individual polarization components are unlikely to be independently configurable. It is further noted that because the virtualized Willis metamaterial can also handle complex polarizabilities, it is possible to achieve controllable gain and loss of the system as well as complex bianisotropy, enabling phenomena such as imaginary reciprocal and nonreciprocal coefficients which are impossible with conventional bianisotropic media[29].

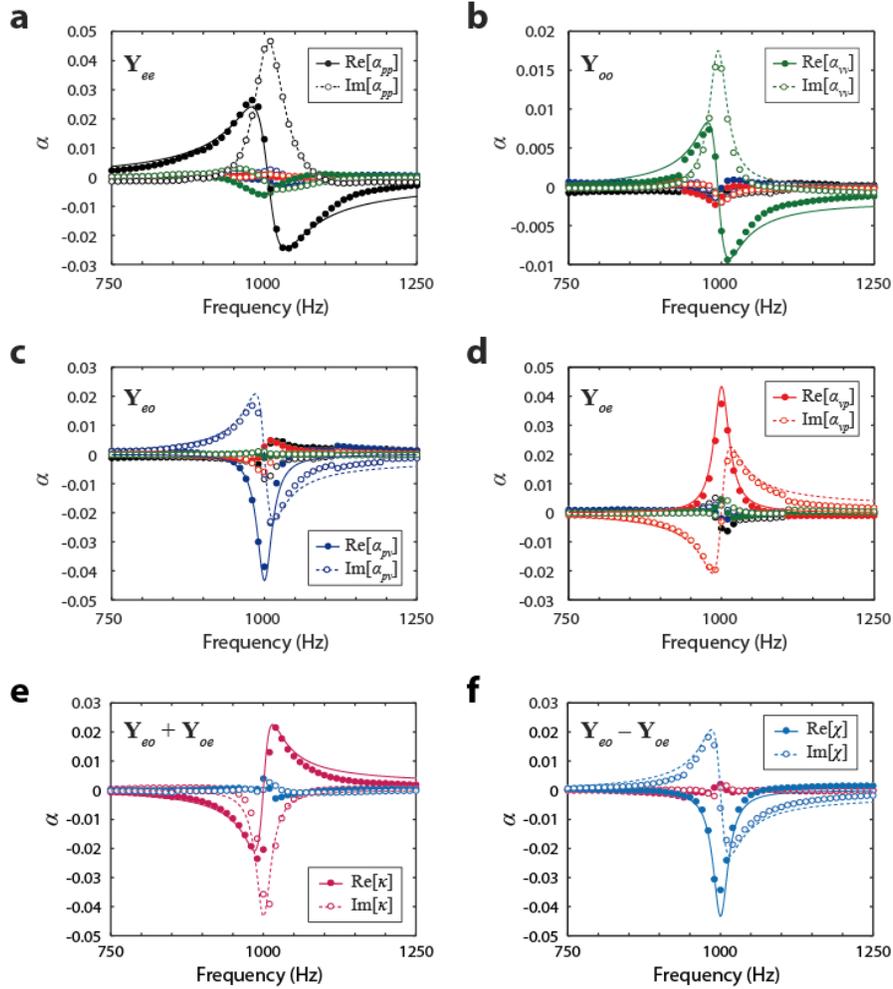

**Figure 3 | Decoupled excitation of polarization components. a-d** Virtualized Willis metamaterial for the four basis convolution matrices $e_{ee}$ (**a**), $e_{oo}$ (**b**), $e_{eo}$ (**c**), and $e_{oe}$ (**d**) with the same Lorentzian convolution kernel $Y_0$, where the model parameters are set to $a = 2\pi \times 15$, $\theta = -\pi/2$, $\gamma = 2\pi \times 15$ Hz, and $\omega_0 = 2\pi \times 1.0$ kHz. The polarizabilities $\alpha_{pp}$, $\alpha_{vv}$, $\alpha_{pv}$, and $\alpha_{vp}$ are depicted in black, green, blue and red, respectively, with solid/empty symbols representing the real/imaginary parts of the experimental results. The corresponding theoretical models are plotted with solid/dashed lines for the real/imaginary parts. **e, f** Purely reciprocal and purely nonreciprocal Willis couplings $\kappa = i(\alpha_{pv} - \alpha_{vp})/2$ and $\chi = (\alpha_{pv} + \alpha_{vp})/2$, realized with a balanced $e_{oe}$ and $e_{eo}$. The purely reciprocal Willis coupling satisfying $\alpha_{vp} = -\alpha_{pv}$ is demonstrated by their summation, i.e., $Y = Y_0(e_{eo} + e_{oe})$ (**e**), and the purely nonreciprocal term satisfying $\alpha_{vp} = \alpha_{pv}$ is demonstrated by subtracting the two basis convolution matrices, $Y = Y_0(e_{eo} - e_{oe})$ (**f**).

**Willis coupling beyond the passivity bound** The ability to excite Willis coupling is known to be limited by the passivity condition, as discussed earlier. Following the derivation in Ref. 25 for two- and three-dimensional systems, for the one-dimensional passive system treated here, the maximum bianisotropy bound is dictated by the following two inequalities (see Supplementary Note S3 for the derivation and a discussion of the maximum bound):

$$\begin{aligned}|s_{oe}|^2 + |1+s_{ee}|^2 \leq 1, \\ |s_{eo}|^2 + |1+s_{oo}|^2 \leq 1.\end{aligned} \quad (8)$$

or, equivalently in terms of polarizability,

$$\begin{aligned}|k_0 \alpha_{vp}|^2 + |1+ik_0 \alpha_{pp}|^2 \leq 1, \\ |k_0 \alpha_{pv}|^2 + |1+ik_0 \alpha_{vv}|^2 \leq 1.\end{aligned} \quad (9)$$

Thus, the maximum bound of the Willis coupling is given by $|\alpha_{vp}|$ ($|\alpha_{pv}|$) $\leq k_0^{-1}$, where the equality is satisfied when $\alpha_{pp}$ ($\alpha_{vv}$) = $ik_0^{-1}$. If the systems are strictly reciprocal, e.g., in the case of a physically designed structure with curled channels[25], we can set $t_+ = t_-$, and then, the inequality is reduced to $|r_+ - r_-| \leq 2$ as an upper bound on the Willis coupling term for a reciprocal scatterer. In this passive case, one needs to design a system with $t_+ = t_- = 0$ and $r_+ = -r_- = e^{i\varphi}$, where $\varphi$ is the arbitrary real number needed to approach the equality in the inequality. However, with our implementation using a virtualized metamaterial, we need not be restricted by the reciprocity and maximum bound of the Willis coupling. Because the secondary radiation source in our virtualized metamaterial draws power from external digital circuits, it becomes straightforward to overcome the maximum bound of the Willis coupling.

Figure 4 shows the magnitudes of the Willis couplings $|\alpha_{pv}|$ (Fig. 4a) and $|\alpha_{vp}|$ (Fig. 4b) for Lorentzian convolution kernels with two different scaling factors ($a = 2\pi \times 15$ and $2\pi \times 30$) at different resonance frequencies ($f_0 = 900$ Hz, 1000 Hz, and 1100 Hz) and with a fixed bandwidth ($\gamma = 2\pi \times 15$ Hz). All cases show a central peak at the resonance frequency where the Willis coupling is maximized, in agreement with the analytical models in Eq. (6). In contrast to the maximum bianisotropy of a passive metamaterial (black dashed line), which is dictated by $|\alpha_{vp}|$ (or $|\alpha_{pv}|$) = $k_0^{-1}$, the newly established maximum bianisotropy for the virtualized Willis metamaterial (magenta dotted line) is now modified to $|\alpha_{vp}|$ (or $|\alpha_{pv}|$) = $2k_0^{-1}|\sin(k_0\Delta)Y_0|$ with $|Y_0| \cong a\gamma^{-1}/2$ at the resonance frequency, revealing the set of parameters for controlling the Willis coupling strength (Supplementary Fig. S3). It is worth mentioning that the presence of $k_0\Delta$ in $\sin(k_0\Delta)$ reveals the required metamaterial geometry of the scatterer (or source) layout, while the amplitude $a$, bandwidth $\gamma$, and center frequency $\omega_0$ reveal the significance of the scatterer characteristics. At the values $\Delta = 50$ mm and $\gamma = 2\pi \times 15$ Hz used in the experiment, Fig. 4 shows theoretical (lines) and experimentally realized (symbols) Willis parameters with the scaling factor of $a = 2\pi \times 15$ and $a = 2\pi \times 30$ respectively, each for below and above the passivity bound (black-dashed lines). In addition to the control parameter $a$, which represents the power drawn by the active devices, it is further noted that the layout of the scatterers, represented by $\Delta$, can also be used to control the strength of the polarizability. In our implementation, a small $\Delta \sim \lambda/7$ was used, in the regime of metamaterials without further optimization. With the introduction of a resonance directly into the Willis coupling term, the system will draw the necessary power from the external source, i.e., become active, and the conventional Willis bound can intuitively be surpassed.

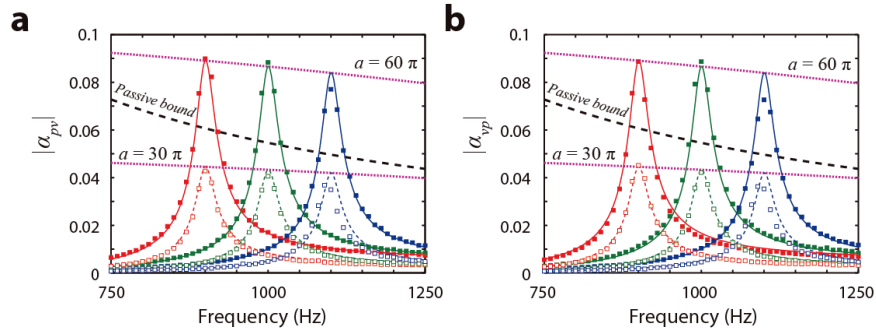

**Figure 4 | Willis coupling beyond the passivity bound.** Willis coupling beyond the passivity bound achieved by controlling the scaling factors. **a**, The magnitude of $\alpha_{pv}$ for the odd-to-even convolution kernel $\mathbf{e}_{eo}$. The Lorentzian responses at three center frequencies, $f_0 = 900$ Hz (red), 1000 Hz (green) and 1100 Hz (blue), are demonstrated with two different scaling factors, $a = 2\pi \times 15$ (empty symbols) and $2\pi \times 30$ (filled symbols). The analytical results for each scaling factor are also plotted as solid and dashed lines, and the magenta dotted lines denote the theoretical values of the Lorentzian peaks at the resonance frequencies. The black dashed line represents the passivity limit of Willis coupling, i.e., $|\alpha_{pv}| = k_0^{-1}$. **b**, Same as **a** except that $|\alpha_{vp}|$ for the even-to-odd convolution kernel $\mathbf{e}_{oe}$ is demonstrated.

**Inverse design of the broadband Willis response** Recalling that there is no reason for the frequency response of $Y$ to be limited to a Lorentzian response in our implementation, here, we address a metamaterial realization with an arbitrary target response function $F_0(\omega)$ based on the notion of inverse design. To realize $\alpha_{pv}(\omega)$ (or $\alpha_{vp}(\omega)$) = $F_0(\omega)$, we utilize the relation in Eq. (6) and obtain the convolution function $Y_{eo}(\omega) = \frac{1}{2} F_0(\omega) k_0 \sin(k_0 \Delta)^{-1}$ (or $Y_{oe}(\omega) = -\frac{1}{2} F_0(\omega) k_0 \sin(k_0 \Delta)^{-1}$) for the target frequency response $F_0$. By applying inverse Fourier transformation to $Y_{eo}(\omega)$ (or $Y_{oe}(\omega)$), we can numerically obtain the required time-domain convolution function $y(t)$. As metamaterials restrict $k_0 \Delta$ to be small, the resultant time-domain function of this inverse design process will be similar to the inverse Fourier transform of the original target frequency response $F_0(\omega)$. For example, we consider an intriguing target frequency spectrum with a flat broadband response between $\omega_1$ and $\omega_2$, specifically,

$$F_0(\omega) = \left[-u(|\omega|-\omega_1) + u(|\omega|-\omega_2)\right] + \frac{i}{\pi}\log\left|\frac{(\omega+\omega_1)(\omega-\omega_2)}{(\omega-\omega_1)(\omega+\omega_2)}\right|, \tag{10}$$

which satisfies the Kramers-Kronig (KK) relation, along with its inverse Fourier transform,

$$f_0(t) = \frac{2}{\pi}\frac{a}{t}[\sin(\omega_1 t + \theta) - \sin(\omega_2 t + \theta)]u(t). \tag{11}$$

When $\omega_2$ is set to be much larger than $\omega_1$, the above function $F_0(\omega)$ with $\theta = 0$ ($\theta = \pi/2$) provides a flat real (imaginary) spectrum over a broad frequency range while suppressing the imaginary (real) part, while peaks appear in the vicinity of $\omega_1$ and $\omega_2$. This $F_0(\omega)$ could thus be used to design broadband Willis metamaterials offering purely real or purely imaginary polarizability.

To demonstrate the purely reciprocal and nonreciprocal Willis couplings, also with broadband characteristics, we then program the convolution kernel to be $\mathbf{Y} = Y_0(\mathbf{e}_{eo} + \mathbf{e}_{oe})$ for the reciprocal case and $\mathbf{Y} = Y_0(\mathbf{e}_{eo} - \mathbf{e}_{oe})$ for the nonreciprocal case, as used in the narrowband demonstrations shown in Fig. 3e and 3f. Figure 5 shows the experimental realization of the purely reciprocal Willis parameter $\kappa = i(\alpha_{pv} - \alpha_{vp})/2$ and the purely nonreciprocal parameter $\chi = (\alpha_{pv} + \alpha_{vp})/2$, achieving a flat broadband spectrum over $(\omega_1, \omega_2) = (800 \text{ Hz}, 1200 \text{ Hz})$ for $\theta = 0$ (Fig. 5b,c) and $\theta = \pi/2$ (Fig. 5a,d), with $a = 0.225$. It is important to note that while Fig. 5a and Fig. 5b each correspond to conventional Willis couplings for omega media and moving media, which have real components of $\kappa$ and $\chi$, respectively, the Willis couplings shown in Fig. 5c,d newly achieve imaginary $\kappa$ and $\chi$ values, providing an additional degree of freedom in terms of energy, i.e., gain and loss in the Willis coupling. We note that while even more general frequency responses can be constructed beyond the Lorentzian resonance and flat dispersion demonstrated here, it is necessary to keep some reservations due to the causality restriction. For example, the required time-domain convolution function from the inverse Fourier transform of the target frequency response could contain anticausal components, i.e., $y(t) \neq 0$ for $t < 0$ (see Supplementary Note S4 for the flat broadband dispersion of the inverse bulk modulus and mass density). Nonetheless, we are open to the possibility of mitigating at least the condition of $y(t < 0) \neq 0$ through some modification of the virtual metamaterial configuration, such as placing the microphones before the speakers. In essence, our virtualization scheme provides a one-step implementation, through the digitization of the impulse response as a software entity, to obtain any physically allowed broadband spectrum. The same approach can be readily applied in other applications requiring a broadband response. For example, it can also be applied to design causality-optimal sound absorption media[30], with the advantage that once the causality-optimal spectrum has been formulated, there is no need to further formulate a strategy to obtain the corresponding metamaterial structures. In other words, by adopting the virtualization approach, one does not need to be concerned with passivity and reciprocity which are the usual starting points for formulating performance bounds[31], but rather can relax the necessary considerations to causality only. In our case, causality is considered automatically through the KK relation.

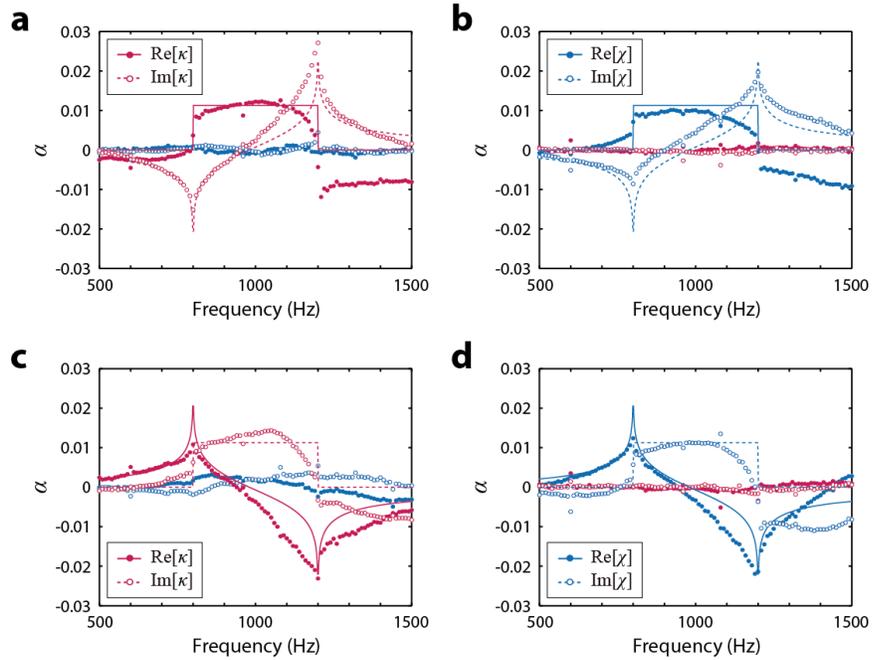

**Figure 5 | Broadband frequency dispersion control of Willis couplings.** Experimental demonstration of Willis coupling parameters inversely designed from a frequency dispersion response with broadband flat real values, while imaginary values are restricted in the vicinity of the band edges. By balancing $\alpha_{pv}$ and $\alpha_{vp}$, (**a, c**) broadband purely reciprocal Willis coupling parameters $\kappa = i(\alpha_{pv} - \alpha_{vp})/2$ and (**b, d**) broadband purely nonreciprocal Willis coupling parameters $\chi = (\alpha_{pv} + \alpha_{vp})/2$ are achieved. **a** and **b** correspond to conventional omega media and moving media with real $\kappa$ and $\chi$, while **c** and **d** show imaginary reciprocal and nonreciprocal Willis couplings, which are not naturally achievable.

**Discussion**

We demonstrate active bianisotropic metamaterials offering Willis coupling beyond the passivity limit. The conditions for maximum Willis coupling and reciprocity in the passivity regime are revisited, and then, the new bound of the Willis parameter and the reciprocity with the introduction of an active metamaterial are analyzed. By employing a virtualized metamaterial platform that enables the flexible design of scattering properties by means of software convolution functions, Lorentzian resonances of all of the polarizability parameters are demonstrated, with exclusive access to each polarizability, including the Willis parameters, inverse bulk modulus and mass density, independently. Using the fully independent excitation of each parameter as well as precise balancing between them, the operations of purely reciprocal and nonreciprocal Willis couplings are realized. We also demonstrate the breaking of the Willis bound in the passivity limit for the first time while isolating the control parameters involved with the newly established Willis bound in the active regime, such as the amplitude, bandwidth, and frequency of the active resonator that feeds in external power for the scattered fields. Finally, we achieve the inverse design of target Willis responses with identical, flat-amplitude Willis coupling strengths over a broad frequency range, for the reciprocal and nonreciprocal cases as well as the newly revealed case of nonconserved bianisotropy. Demonstrating full control and top-down tailoring of reciprocity, bianisotropy, inverse bulk modulus and mass density dispersion within the same platform, our work achieves the full potential of Willis metamaterials and their diverse applications beyond the passivity limit.

## Methods

**Experimental setup** For the measurement, we used the 4-point measurement method with a National Instruments DAQ device and LabVIEW. The scattering properties of the meta-atom were tested by means of incident waves coming from the forward and backward directions to identify all four scattering parameters ($r_+$, $t_+$, $r_-$, and $t_-$). In the experimental setup, we flipped the orientation of the meta-atom, while the waves were always incident from the same end of the impedance tube.

**Fabrication of the meta-atom** The virtualized meta-atom consists of two MEMS microphones (INMP401) and speakers (SMT-1028-t-2-r) laterally located on each edge of the acrylic frame, which are connected to an external single-board computer (Raspberry Pi 4B+) with amplifiers and analog-to-digital/digital-to-analog converters (see Fig. 2). For digital processing, the input signals sampled by the microphones are digitally processed by the microprocessor and then fed to the speakers in real time with a sampling frequency of $f_s = 7.5$ kHz and a number of samples equal to $N = 400$. The convolution is calculated as $S_i^V[n] = \sum_j \sum_{k=0}^{N} Y_{ij}[k](M_j[n-k] - M_j[n-k-1])$, where the index $n = t/T_s$ is the discrete time with sampling period $T_s = f_s^{-1}$. The speakers and microphones, which communicate with the microprocessor through the SPI (Serial Peripheral Interface), are mounted in an acrylic frame (width = 3.0 cm, length = 6.5 cm). This transducer module is, in turn, mounted on the acoustic waveguide (width = 3.0 cm, height = 3.0 cm). The total temporal delay from a microphone to a speaker is 277 μs, corresponding to a 1.74 rad phase lag at 1 kHz.

## Data availability

The data that support the plots in this paper and other findings of this study are available from the corresponding author upon reasonable request.

**Acknowledgments**

J. L. acknowledges funding from the Research Grants Council under Grant No. 16303019. N. P. was supported by the NRF of Korea through the Global Frontier Program (2014M3A6B3063708).

**Author Contributions**

J. L. conceived the idea of the virtualization of a Willis metamaterial. C. C. and J. L. established the maximum Willis coupling in a one-dimensional system. C. C. and N. P. established the inverse design of broadband Willis coupling. C. C. and X. W. established the setup of the atom and the control program and performed the measurements. All authors contributed to the data analysis and the writing of the manuscript. J. L. and N. P. managed the project.

**Additional Information**

Supplementary information is available in the online version of this paper. Reprints and permissions information are available online at http://www.nature.com/reprints. Correspondence and requests for materials should be addressed to J.L. and N.P.

**Competing Financial Interests**

The authors declare no competing interests.